\documentclass[prl,twocolumn,superscriptaddress]{revtex4}

\usepackage{amsfonts,amsmath,amssymb,amsthm}
\usepackage{epic,eepic,graphicx,mathpple}

\usepackage{amsmath, amsthm, amssymb}
\usepackage{graphicx}

\theoremstyle{definition}

\theoremstyle{remark}

\newcommand{\eeee}{\textrm{ee}}
\newcommand{\eeii}{\textrm{ei}}
\newcommand{\iiii}{\textrm{ii}}
\newcommand{\iiee}{\textrm{ie}}
\newcommand{\ee}{\textrm{e}}
\newcommand{\ii}{\textrm{i}}

\begin{document}

\keywords{}

\title{Balanced networks of spiking neurons with spatially dependent recurrent connections}

\author{Robert Rosenbaum}
\affiliation{Department of Mathematics, University of Pittsburgh, Pittsburgh PA 15260, USA}
\affiliation{Center for the Neural Basis of Cognition, Pittsburgh PA 15213, USA}
\author{Brent Doiron}
\affiliation{Department of Mathematics, University of Pittsburgh, Pittsburgh PA 15260, USA}
\affiliation{Center for the Neural Basis of Cognition, Pittsburgh PA 15213, USA}

\begin{abstract}
Networks of model neurons with balanced recurrent excitation and inhibition produce irregular and asynchronous spiking activity.  We extend the analysis of balanced networks to include the known dependence of connection probability on the spatial separation between neurons.  In the continuum limit we derive that stable, balanced firing rate solutions require that the spatial spread of external inputs be broader than that of recurrent excitation, which in turn must be broader than or equal to that of recurrent inhibition.  For finite size networks we investigate the pattern forming dynamics arising when balanced conditions are not satisfied.  The spatiotemporal dynamics of balanced networks offer new challenges in the statistical mechanics of complex systems.       
\end{abstract}

\maketitle

The study of spatiotemporal dynamics and variability in complex systems is at the interface of the physical, chemical, biological, and social sciences~\cite{sagues:2007,lindner:2004}.  
In the neurosciences, a longstanding topic of interest is the significant variability in cortical neuron spike train responses~\cite{maimon:2009,churchland:2010}.  Models of cortical networks capture this high variability when recurrent excitatory and inhibitory inputs are balanced.  {Such ``balanced networks'' show irregular and asynchronous spiking dynamics through a complex, sometimes chaotic, network state~\cite{vanVreeswijk:1996us}.}  Nevertheless, the statistics of balanced networks are amenable to mean field analysis~\cite{vanVreeswijk:1998uz,Brunel:2000th,Renart:2004tn,boustani:2009,Renart:2010hj}, using techniques developed for spin-glass systems~\cite{binder:1986}.   
Subsequent  experiments in cortex lend support to balanced network states with measurements of large and opposing excitatory and inhibitory synaptic currents~\cite{Shu:2003ht,Haider:2006gs}, asynchronous cortical activity~\cite{Ecker:2010dn}, as well as the sensitivity of network dynamics to small perturbations~\cite{london:2010}. 

The probability that two cortical neurons are connected depends on their separation in physical space or, for some sensory systems, feature space~\cite{holmgren:2003,Oswald:2009kp,Ko:2011gy,Levy:2012dy}.  There has been substantial theoretical work on the the spatiotemporal dynamics of phenomenological macroscale models of cortex~\cite{coombes:2005,bressloff:2012}.  In contrast, theoretical work in balanced networks assumes a spatially homogeneous or discretely clustered topology~\cite{vanVreeswijk:1996us,Renart:2004tn,LitwinKumar:2012go}.  The capacity for pattern formation and spatial filtering in balanced networks with spatially dependent connection probabilities has not been addressed.

In this letter, we derive experimentally testable conditions on the strength and spatial profile of connection probabilities that must be satisfied for a recurrent network of excitatory and inhibitory neuron models to maintain a stable balanced state in the continuum limit.  {Specifically, we find that external inputs must be broader than recurrent excitation, which in turn must be broader than or equal in broadness to recurrent inhibition.}  
Further, we investigate  spatiotemporal spiking dynamics when stable balanced solutions do not exist.

\paragraph{Network model.}

We consider a network of $N$ integrate--and--fire neurons, half of which are excitatory and half inhibitory, spaced evenly on the state space $\Gamma=(0,1]$, so that the $k$th excitatory or inhibitory neuron is at location $x=2k/N$.  
The input current to the $k$th excitatory ($\alpha=\ee$) and inhibitory ($\alpha=\ii$) neuron is given by 
\vspace{-0.3cm}
\begin{equation}\label{E:Is}
I_\alpha(x,t)=\sum_{j=1}^{N/2}  J_{\alpha \ee}^{k,j}s_{\ee,j}(t)-J_{\alpha \ii}^{k,j}s_{\ii,j}(t)+J_\alpha(x),
\end{equation}
respectively, where  $x=2k/N$ and 
$s_{\ee,j}(t)=\sum_i \delta(t-t^i_{\ee,j})$
is the spike train of the $j$th excitatory neuron and similarly for $s_{\ii,j}(t)$. 
Static external input is provided by the terms $J_\alpha(x)$. 
The synaptic weight, $J_{\alpha \beta}^{kj}$, is equal to the constant $J_{\alpha \beta}$ with probability $k^\Gamma_{\alpha\beta}(x-y)$, else it is zero ($\alpha,\beta \in\{\ee,\ii\}$).
Here, $k^\Gamma_{\alpha\beta}(x)=\sum_{n=-\infty}^\infty k_{\alpha\beta}(x+n)$ so that $\Gamma$ has periodic boundaries and $k_{\alpha\beta}$ is the spatial profile of $\beta$ to $\alpha$ connectivity.  As in~\cite{vanVreeswijk:1996us,vanVreeswijk:1998uz}, we fix $k_{\alpha\beta}(x)\ll 1$ to assure asynchrony and we then consider the behavior of the network as $N\to\infty$.  

Cortical neurons receive a large number of high amplitude excitatory inputs, implying that a post-synaptic cell only requires only a fraction of excitatory pre-synaptic cells to drive spike responses~\cite{shadlen:98}.  Following past studies in balanced networks~\cite{vanVreeswijk:1996us,vanVreeswijk:1998uz,Renart:2010hj} we model this with an $\mathcal O(1)$ distance between rest and spike threshold and consider $J_{\alpha \beta}\sim \mathcal O(1/\sqrt N)$,  $k_{\alpha\beta}\sim \mathcal O(1)$ and $J_\alpha(x)\sim \mathcal O(\sqrt N)$.    
To simplify calculations we define $j_{\alpha \beta}=J_{\alpha\beta}\sqrt N$,  $j_\alpha(x)=J_\alpha(x)/ \sqrt N$ which do not depend on $N$.

Under these scaling assumptions, a neuron receives recurrent input from $\mathcal O(N)$ excitatory neurons but only requires $\mathcal O(\sqrt N)$ excitatory inputs to be active in an integration window to produce a spike.  Finite firing rates are therefore only maintained in the continuum limit through a dynamically stable balance between excitation and inhibition \cite{vanVreeswijk:1996us,vanVreeswijk:1998uz,Renart:2010hj}.  We next derive conditions under which such a stable balanced network state exists.

\paragraph{Conditions on the existence of a balanced state in the continuum limit.}

The mean firing rates of neurons in the network are denoted by 
$
\nu_\alpha(x)=\langle E[ s_{\alpha,k}(t)]\rangle$, where $E[\cdot]$ represents expectation over network connectivity and $\langle \cdot \rangle$ the average over time.
In the continuum limit,  the mean input currents are related to the firing rates by 
\begin{equation}\label{E:mu1}
\begin{aligned}
\mu_\alpha(x)&:=\langle E[I_\alpha(x,t)]\rangle\\
&=\sqrt{N}\left[   w_{\alpha\ee}*\nu_\ee(x) -  w_{\alpha \ii}* \nu_\ii(x)+j_\alpha(x) \right]
\end{aligned}
\end{equation}
for $\alpha=\ee,\ii$ where $w_{\alpha\beta}(x)=j_{\alpha\beta}k_{\alpha\beta}(x)$ and $*$ denotes circular convolution on $\Gamma$.
Similarly, {the infinitesimal temporal variances of the input currents are given by 
\begin{align}
V_\alpha(x)&:=\lim_{\delta\to 0}\delta^{-1}\left\langle E\left[\int_t^{t+\delta} I_\alpha(x,s)-\mu_\alpha(x)ds\right]^2\right\rangle
\notag\\
&=j_{\alpha\ee} w_{\alpha\ee}*\nu_\ee(x) +j_{\alpha\ii} w_{\alpha\ii}* \nu_\ii(x).\label{E:var1}
\end{align}
We aim to derive conditions under which $\nu_\alpha(x)$, $\mu_\alpha(x)$ and $V_\alpha(x)$ each converge to a finite limit as $N\to\infty$ and $\nu_\alpha(x)$ does not become identically zero.  For these conditions to be realized, we must have that
\begin{equation}\label{E:mubalance}
w_{\alpha\ee}*\nu_\ee(x) -  w_{\alpha \ii}* \nu_\ii(x)+j_\alpha(x) =\mathcal O(1/\sqrt N).
\end{equation}
Taking $N\to\infty$ gives a Fredholm equation of the first kind whose solution, when it exists, is given in the Fourier domain by
\begin{equation}\label{E:S0}
\begin{aligned}
\widetilde \nu_\ee&=\frac{\widetilde j_\ee \widetilde w_\iiii-\widetilde j_\ii \widetilde w_\eeii}{\widetilde w_\eeii \widetilde w_\iiee-\widetilde w_\eeee
   \widetilde w_\iiii}\\
\widetilde \nu_\ii&=\frac{\widetilde j_\ee \widetilde w_\iiee-\widetilde j_\ii \widetilde w_\eeee}{\widetilde w_\eeii \widetilde w_\iiee-\widetilde w_\eeee
   \widetilde w_\iiii}
\end{aligned}
\end{equation}
where $\widetilde f(n)=\int_\Gamma e^{-2\pi x n i}f^\Gamma(x)dx$.
This equality must hold at every Fourier mode, $n$,  for which $\widetilde w_\eeii(n) \widetilde w_\iiee(n)-\widetilde w_\eeee(n) \widetilde w_\iiii(n)\ne 0$.
If $\widetilde w_\eeii(n) \widetilde w_\iiee(n)-\widetilde w_\eeee(n) \widetilde w_\iiii(n)= 0$ at some Fourier mode, then for a solution to exist, it must also be true that $\widetilde j_\ee(n) \widetilde w_\iiii(n)-\widetilde j_\ii(n) \widetilde w_\eeii(n)=\widetilde j_\ee(n) \widetilde w_\iiee(n)-\widetilde j_\ii(n) \widetilde w_\eeee(n)=0$ at that Fourier mode.

Requiring firing rates to be non-negative and not identically zero implies that 
\begin{align}
\frac{\overline j_\ee}{\overline j_\ii}&>\frac{\overline w_\eeii}{\overline w_\iiii}>\frac{\overline w_\eeee}{\overline w_\iiee}\label{E:balance1}\;\;\textrm{ or}\\
\label{E:balance2}\frac{\overline j_\ee}{\overline j_\ii}&<\frac{\overline w_\eeii}{\overline w_\iiii}<\frac{\overline w_\eeee}{\overline w_\iiee}
\end{align}
where $\overline f=\widetilde f(0)=\int_\Gamma f^\Gamma(x)dx$.
Note that Eq.~\eqref{E:balance1} is equivalent to a balance condition derived in~\cite{vanVreeswijk:1998uz} for spatially homogeneous networks.  We show below that Eq.~\eqref{E:balance1} leads to a stable balanced state for large $N$ but Eq.~\eqref{E:balance2} does not.
The solution in Eq.~\eqref{E:S0} is only viable if $\widetilde \nu_\alpha$ has a well-defined inverse Fourier transform, which requires at least that
\begin{equation}\label{E:ncond}
\lim_{n\to\infty}\frac{\widetilde j_\ee(n) \widetilde w_{\ii\alpha}(n)-\widetilde j_\ii(n) \widetilde w_{\ee \alpha}(n)}{\widetilde w_\eeii(n) \widetilde w_\iiee(n)-\widetilde w_\eeee(n)
   \widetilde w_\iiii(n)}=0
\end{equation}
for $\alpha=\ee,\ii$.  We investigate this condition for specific examples below.

\paragraph{Example with Gaussian connectivity --}

The analysis of the balanced state above is valid in the $N\to\infty$ limit for a large class of neuron models~\cite{Renart:2004tn}. To find solutions at large but finite system size, we use a leaky integrate--and--fire (LIF) model~\footnote{Membrane potentials satisfy $v'(t)=-v(t)/\tau_m+I_{\alpha}(x,t)$ with a reflecting barrier at $v(t)=-1$ where $\tau_m=20$~ms.  Spikes occur whenever $v(t)=1$ at which point  $v(t)$ is reset to zero.\label{E:LIFdef}}.  
Steady-state firing rates can be found numerically using Monte Carlo simulations of the full network or by searching for a fixed point $[\nu_\ee^0(x),\nu_\ii^0(x)]$ that satisfies
\begin{equation}\label{E:nuphi}
\nu_\alpha^0(x)=\phi(\mu_\alpha^0(x),V_\alpha^0(x))
\end{equation}
where $\phi(\mu,V)$ relates input mean and variance to firing rate of the LIF model in the diffusion limit~\footnote{$\phi(\mu,V)$ is known in closed form~\cite{Amit:1997uj} but more efficiently calculated by solving a boundary value problem~\cite{Richardson:2007ct}.} and where $\mu^0_\alpha(x)$ and $V^0_\alpha(x)$ are given in terms of $\nu^0_\ee(x)$ and $\nu^0_\ii(x)$ by Eqs.~\eqref{E:mu1}-\eqref{E:var1}.   {Numerical solutions to Eq.~\eqref{E:nuphi} were used for the curves labeled ``FP'' in Figs.~\ref{F:balance}-\ref{F:NarrowInput}.}

\begin{figure}
\centering{
\includegraphics[width=3.35in]{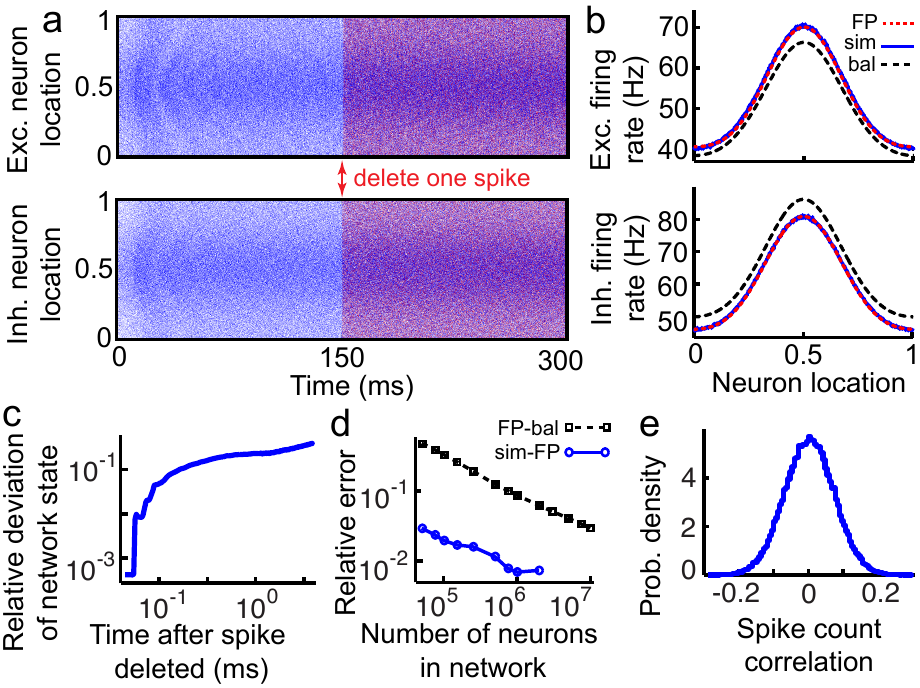}
}
\caption{{\bf Balanced network dynamics.} (Color online) {\bf (a)} Raster plots from two Monte-Carlo simulations with identical initial states (blue and red dots respectively, $N=10^5$).  In one simulation (red dots), the first spike after 150~ms was skipped, revealing a sensitivity to perturbations.   {\bf (b)}  Population firing rates $\nu_e(x)$ and $\nu_i(x)$ calculated from full network simulations (solid blue line, $N=2\times10^6$), the fixed point from Eq.~\eqref{E:nuphi} (dotted red line, $N=2\times10^6$), and the balanced solution from Eq.~\eqref{E:GS0} (dashed black line).
{\bf (c)} The normalized $L^2$ deviation between the vector of membrane potentials of the two network simulations from (a) 
as a function of time elapsed since a spike was skipped.    {\bf (d)}  Relative $L^2$ distance between the balanced state and the fixed point (dashed black line with squares), and between the fixed point and network simulations (solid blue line with circles). {\bf (e)} Histogram of spike count correlations between neighboring neurons ($N=10^5$; mean of $4.84\times 10^{-4}$ and a standard deviation of 0.0711).  Correlations computed by counting spikes over a 300~ms window in each neuron from 200 simulations of the same network realization with different initial conditions.
}
\label{F:balance}
\end{figure}

For ease of exposition, we consider Gaussian shaped connectivity kernels and we assume that probability (but not strength) of a connection depends only on presynaptic cell type.  In particular, we set $w_{\alpha\beta}(x)=\overline w_{\alpha\beta}w_\beta(x)$ and $j_\alpha(x)=\overline j_\alpha j(x)$ where 
$$
w_\beta(x)=\frac{1}{\sqrt{2\pi}\sigma_\beta}e^\frac{-x^2}{2\sigma_\beta^2},
\hspace{.1in}
j(x)=1-p+\frac{p}{\sqrt{2\pi}\sigma_o}e^\frac{-(x-x_o)^2}{2\sigma_o^2},
$$
satisfy $\overline j=\overline w_\beta=1$ for $\alpha,\beta\in\{\ee,\ii\}$.  In this case, the balance condition in Eq.~\eqref{E:ncond} is satisfied only if $\sigma_o> \sigma_\ee,\sigma_\ii$.  Hence, external inputs must be spatially broader than recurrent connections for a balanced solution to exist.  Under this condition, taking the inverse transform in Eq.~\eqref{E:S0} gives the balanced solutions
\begin{equation}\label{E:GS0}
\nu_\alpha(x)=\overline \nu_\alpha \left[\frac{p}{\sqrt{2\pi(\sigma_o^2-\sigma_\alpha^2)}}e^{-\frac{-(x-x_o)^2}{2(\sigma_o^2-\sigma_\alpha^2)}}+(1-p)\right]
\end{equation}
where $\overline \nu_\alpha=\widetilde \nu_\alpha(0)$ from Eq.~\eqref{E:S0}.  Note that the peaked shape of the firing rate profile from Eq.~\eqref{E:GS0}, though spatially filtered by recurrent activity, is inherited by the peaked shape of the inputs.  Flat inputs ($p=0$) lead to a flat firing rate profile ($\nu_\alpha(x)=\overline \nu_\alpha$).

When the balanced state exists~\footnote{
Unless otherwise specified, parameters for all simulations are $\sigma_\ee=\sigma_\ii=0.1$, $\sigma_o=0.2$, $j_\eeee=0.5$, $j_\eeii=1$, $j_\iiee=0.7$, $j_\iiii=1$, $j_\ee=4\times 10^{-4}$, $j_\ii=3\times 10^{-4}$, $p=0.25$, $\overline k_{\alpha\beta}=0.02$ for $\alpha,\beta\in\{\ee,\ii\}$.}, simulations show asynchronous and irregular spiking dynamics (Fig.~\ref{F:balance}a).  {The microscopic state of the network is highly sensitive to the deletion of a single spike (Fig.~\ref{F:balance}a,c), but sufficiently small perturbations of the membrane potentials do not cause a divergence of trajectories (not pictured).  These findings are consistent with previous studies showing that balanced networks can exhibit ``stable chaos'' characterized by exponentially long transients and insensitivity to sufficiently small perturbations~\cite{Politi:1993,Ginelli:2002,Vogels:2005,boustani:2009,Monteforte:2012hh}.}

The macroscopic dynamics, measured by the network firing rates, is stable to the deletion of spikes.  The firing rates are given by fixed point of Eq.~\eqref{E:nuphi}, which converges to the balanced fixed point given by Eq.~\eqref{E:GS0} as the network size increases (Fig.~\ref{F:balance}b,d).  The distribution of Pearson correlation coefficients between the spike counts of neighboring neurons is approximately Gaussian-shaped with a mean near zero despite the fact that neighboring neurons share more than 5\% of their inputs (Fig.~\ref{F:balance}e), consistent with the network having reached a stable asynchronous state~\cite{Renart:2010hj}.

\begin{figure}
\centering{
\includegraphics[width=3.35in]{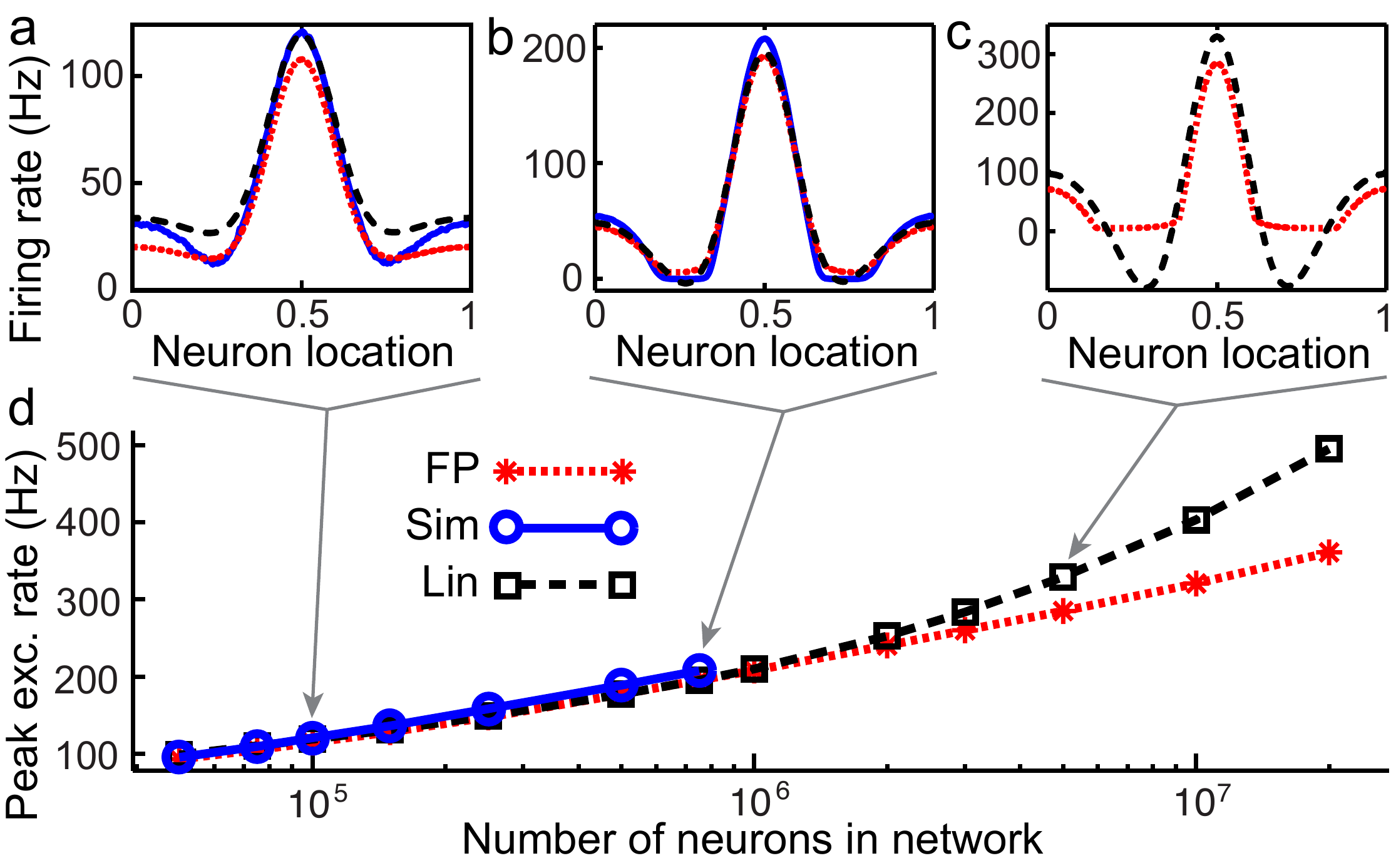}
}
\caption{{\bf Loss of balanced state if external inputs are narrower than recurrent connections.} (Color online) Firing rates of the excitatory population when external inputs are narrower than recurrent inputs for system sizes {\bf (a)} $N=10^5$, {\bf (b)} $N=7.5\times 10^5$, and {\bf (c)} $N=5\times 10^6$.  {\bf (d)}  Peak firing rate of the excitatory population as a function of system size.  In all panels $\sigma_o=0.1$, $\sigma_\ee=\sigma_\ii=0.2$, and other parameters are as in Fig.~\ref{F:balance}.  Dotted red curves computed by numerically solving Eq.~\eqref{E:nuphi}. Solid blue curves computed from full network simulations. Dashed black line computed from the linear approximation given in Eq.~\eqref{E:S2}.
}
\label{F:NarrowInput}
\end{figure}

\paragraph{Spatially imbalanced networks --}
An $\mathcal O(c)$ deviation of the firing rates away from balance yields an $\mathcal O(c\sqrt N)$ deviation of the mean input currents, but only  an $\mathcal O(c)$ perturbation of the input variance, \emph{c.f.} Eqs.~\eqref{E:mu1}--\eqref{E:var1}.  When mean input is large in magnitude, the firing rate transfer of an LIF neuron can be approximated as threshold-linear, motivating the following mean-field approximation to firing rate dynamics,
\begin{equation}\label{E:LinApprox}
\tau_m\frac{\partial \nu_\alpha}{\partial t}=-\nu_\alpha+ \gamma \mu_\alpha \Theta(\mu_\alpha).
\end{equation}
Here $\Theta(\cdot)$ is the Heaviside function, $\tau_m$ is the characteristic timescale of the neurons, $\gamma>0$ is the gain of the neuron~\footnote{We use $\gamma=1$ which is valid for the LIF described above when mean input is large.} 
 and $\mu_\alpha$ is related to $\nu_\beta$ through Eq.~\eqref{E:mu1} for $\alpha,\beta\in\{\ee,\ii\}$.  Eq. ~\eqref{E:LinApprox} can be solved for arbitrary $N$ and will provide intuition for network solutions when condition Eq.~\eqref{E:ncond} is violated. 

If Eq.~\eqref{E:LinApprox} admits a fixed point with strictly positive firing rates, it is given in the Fourier domain by
\begin{equation}\label{E:S2}
\begin{aligned}
\widetilde \nu_\ee^0&=\frac{\epsilon \widetilde j_\ee+\widetilde j_\ee \widetilde w_\iiii-\widetilde j_\ii \widetilde w_\eeii}{\epsilon^2 -\epsilon \widetilde w_\eeee +\epsilon \widetilde w_\iiii +\widetilde w_\eeii \widetilde w_\iiee - \widetilde w_\eeee \widetilde w_\iiii}\\
\widetilde \nu_\ii^0&=\frac{\epsilon \widetilde j_\ii+\widetilde j_\ee \widetilde w_\iiee-\widetilde j_\ii \widetilde w_\eeee}{\epsilon^2 -\epsilon \widetilde w_\eeee +\epsilon \widetilde w_\iiii +\widetilde w_\eeii \widetilde w_\iiee - \widetilde w_\eeee \widetilde w_\iiii}
\end{aligned}
\end{equation}
where $\epsilon=1/(\gamma\sqrt N)$.
If Eq.~\eqref{E:ncond} is satisfied then the fixed point in Eq.~\eqref{E:S2} converges to the balanced solution in Eq.~\eqref{E:S0} as $N\to\infty$.   If Eq.~\eqref{E:ncond} is violated ($\sigma_o < \sigma_\ee,\sigma_\ii$) then the higher spatial Fourier modes, and therefore peak firing rates, from Eq.~\eqref{E:S2} diverge as $N\to\infty$ (Fig.~\ref{F:NarrowInput}).  
Eventually this growth of higher Fourier modes causes $\nu_\alpha(x) < 0$ for some $x$ (Fig.~\ref{F:NarrowInput}a-c), at which point Eq.~\eqref{E:S2} no longer reflects a fixed point solution to Eq.~\eqref{E:LinApprox}.

\paragraph{Stability of the balanced state --}

The balanced fixed point from Eq.~\eqref{E:S2} is stable for the mean-field model in Eq.~\eqref{E:LinApprox} whenever
\begin{equation}\label{E:A}
A(n)=\left[\begin{array}{cc}-\epsilon+\widetilde w_\eeee(n) & -\widetilde w_\eeii(n)\\
\widetilde w_\iiee(n) & -\epsilon - \widetilde w_\iiii(n)\end{array}\right]
\end{equation}
has eigenvalues with negative real part or, equivalently, when
\begin{equation}\label{E:Stability}
\begin{aligned}
&\widetilde w_\eeii \widetilde w_\iiee-\widetilde w_\eeee \widetilde w_\iiii>\epsilon (\widetilde w_\eeee-\widetilde w_\iiii)-\epsilon^2\\
\textrm{and}\quad&\widetilde w_\eeee-\widetilde w_\iiii<2\epsilon
\end{aligned}
\end{equation}
at each Fourier mode, $n$. For the Gaussian-shaped kernels described above, stability of the balanced state as $N\to\infty$ under this approximation requires that $\overline w_\eeee<\overline w_\iiii$ and $\sigma_\ee\ge \sigma_\ii$ are satisfied in addition to Eq.~\eqref{E:balance1}, but networks satisfying Eq.~\eqref{E:balance2} do not satisfy Eqs.~\eqref{E:Stability} for large $N$.  
The mean-field model predicts instabilities of the balanced state for full network simulations reasonably well (Fig.~\ref{F:waves}).  In particular, when $\sigma_\ee$ is sufficiently smaller than $\sigma_\ii$, $A(n)$ has eigenvalues with positive real part and the balanced fixed point loses stability and different spatial pattern is produced (Fig.~\ref{F:waves}a,c,e).

Further, for $\sigma_\ee/\sigma_\ii$ near the stability transition, network exhibits waves of activity, but the time-averaged firing rates remain close to the balanced fixed point (Fig.~\ref{F:waves}b,d,e). 
The direction that these waves travel depends on initial conditions even when the network remains fixed (data not shown), suggesting a symmetry-breaking multistability.
This spatially coherent activity is not captured by the mean field model in Eq.~\eqref{E:LinApprox} and its theoretical description is outside the scope of this study.  Regardless, our analysis of the mean field approximation  provides a useful explanation for why the balanced state becomes unstable when excitatory projections are too much narrower than inhibitory projections.

 \begin{figure}
 \centering{
 \includegraphics[width=3.35in]{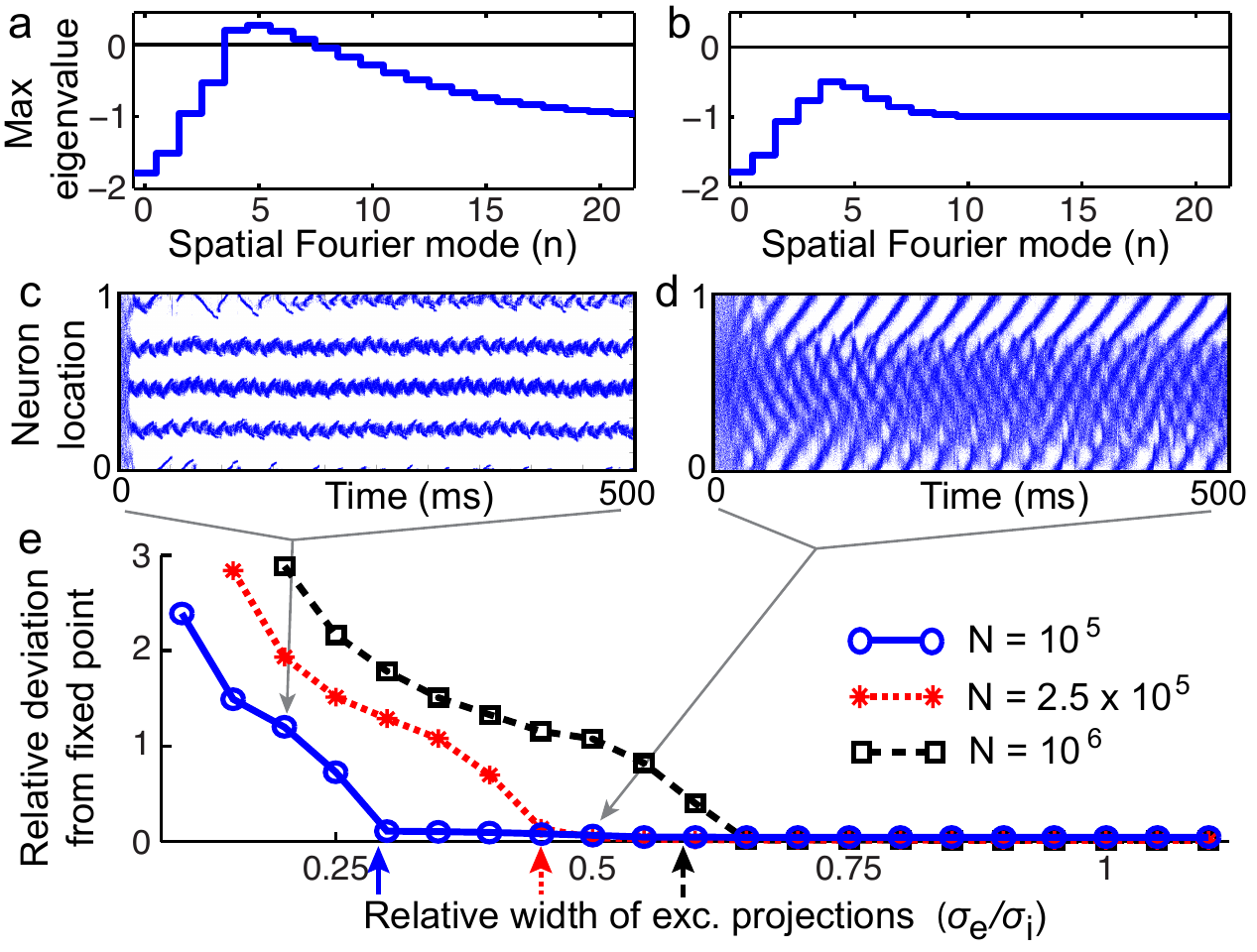}
 }
 \caption{{\bf The balanced state is unstable if recurrent excitation is too narrow compared to inhibition.}  (Color online)  
 {\bf a,b)} Maximum eigenvalue of the matrix $A(n)$ from Eq.~\eqref{E:A} as a function of $n$ with $\sigma_\ee=0.02$ in (a) and $\sigma_\ee=0.05$ in (b) (other parameters as in Fig.~\ref{F:balance}a).  {\bf c,d)} Spike rasters from simulations of the LIF network. {\bf e)} Relative $L^2$ deviation of simulated firing rates from the fixed point determined by Eq.~\eqref{E:nuphi} for various values of $\sigma_\ee$ and $N$ (see inset).   Arrows along horizontal axis mark the smallest value of $\sigma_\ee/\sigma_\ii$ at which some eigenvalue of $A(n)$ has positive real part.
}
\label{F:waves}
\end{figure}

\paragraph{Discussion --}
By taking into account the spatial dependence of connection probabilities, we have derived new conditions for the existence and stability of   balanced solutions.  With Gaussian connectivity, the conditions are simply $\sigma_o > \sigma_\ee \ge \sigma_\ii$.  
Consistent with this conclusion, several studies have found that thalamocortical projections are generally broader than intracortical projections~\cite{landry:1981,freund:1989,rausell:1995} and circuit measurements in cortical layer 4  show that excitation projects more broadly than inhibition~\cite{Levy:2012dy}.  In contrast, many previous models rely on broad inhibition to  sharpen tuning curves~\cite{BenYishai:1995tr,Shapley:2003uj} and promote pattern formation~\cite{coombes:2005,bressloff:2012}.  Our results refute the notion that dynamical mechanisms relying on such broad inhibition can coexist with a balanced state in the continuum limit.  Nevertheless, Eq.~\eqref{E:GS0} reveals that recurrent connections in our model sharpen tuning curves even when $\sigma_\ee\ge\sigma_\ii$.

{For simplicity, we used a one-dimensional single-layer model with periodic boundary conditions.  Our methods can easily be adapted to different spatial topologies.   The analysis of a balanced network on the entire real line is identical to that given in Eqs.~\eqref{E:mu1}--\eqref{E:GS0} except that a continuous Fourier transform takes the place of the discrete transform.  Similarly, if a two-dimensional network is considered, an identical analysis with a two-dimensional Fourier transform yields analogous results.  Our model should be interpreted as a model of a single cortical layer where input from other layers is accounted for by the external inputs, $J_\alpha(x)$.  Recurrent connections between layers can be represented explicitly by adding additional excitatory and inhibitory populations~\cite{folias:2011}, suggesting a possible direction for future work.}

Spatially extended stochastic neural field models are typically constructed by appending additive noise to a deterministic model~\cite{bressloff:2009,bressloff:2012}, similar to the practice of augmenting reaction diffusion systems with additive or multiplicative noise~\cite{sagues:2007}.  Analysis of neural field models driven by external stochastic forcing shows that the spatiotemporal structure of noise is a critical determinant of the ensuing stochastic dynamics~\cite{bressloff2:2012,kilpatrick:2013,hutt:2008}.  In balanced networks, variability arises naturally through internal mechanisms~\cite{vanVreeswijk:1996us,vanVreeswijk:1998uz,Renart:2004tn,LitwinKumar:2012go}, so that assumptions about the structure of external stochastic forcing are not required.  Thus, whereas the study of spatially distributed systems with external stochastic forcing show how pattern forming systems filter noise, balanced networks with spatial interactions offer an alternative framework where complex internal dynamics is the source, as opposed to filter, of spatiotemporal variability.  Our work lays a theoretical foundation for studying such networks and shows that they can exhibit rich dynamics, suggesting several directions for future study.

\section{Acknowledgements}
This work was supported by NIH-1R01NS070865-01A1 and NSF-DMS-1313225.  We thank Ashok Litwin-Kumar, Zachary Kilpatrick, Bard Ermentrout and Jonathan Rubin for helpful discussions.

%

\end{document}